\documentclass[aps,pre,amsmath,amssymb,superscriptaddress,floatfix,showpacs,preprint]{revtex4}
\usepackage{amsmath, amsthm, amssymb, graphicx}
\usepackage{bm}
\input epsf.sty

\begin{document}

\def\g{\gamma}
\def\r{\rho}
\def\av#1{\langle#1\rangle}
\def\pf{P_{\rm F}}
\def\pr{P_{\rm R}}
\def\F#1{{\cal F}\left[#1\right]}

\title{Realm of Validity of the Crooks Relation}

\author{Daniel ben-Avraham}
\affiliation{Department of Physics, Clarkson University, Potsdam, NY 13699-5820, USA}
\affiliation{Department of Mathematics, Clarkson University, Potsdam, NY 13699-5815, USA}

\author{Sven Dorosz}
\affiliation{Theory of Soft Condensed Matter, Universit\'{e} du Luxembourg, Luxembourg, L-1511  Luxembourg}

\author{Michel Pleimling}
\affiliation{Department of Physics, Virginia Polytechnic Institute and State University, Blacksburg, Virginia 24061-0435, USA}

\date{\today}

\begin{abstract}
We consider the distribution $P(\phi)$ of the Hatano-Sasa entropy, $\phi$,  in reversible and irreversible processes,
finding that the Crooks relation for the ratio of the pdf's of the forward and backward processes, $P_F(\phi)/P_R(-\phi)=e^{\phi}$, is satisfied  not only for reversible, but also for irreversible processes, in general, in the adiabatic limit  of ``slow processes."
Focusing on systems with a finite set of discrete states (and no absorbing states), we observe that two-state systems 
always fulfill detailed balance, and obey Crooks relation.  We also identify a wide class of systems, with more than two
states, that can be ``coarse-grained" into two-state systems and obey Crooks relation despite their irreversibility and
violation of detailed balance. We verify these results in selected cases numerically.
\end{abstract}
   
\maketitle
\pagestyle{plain}

\section{Introduction}

In recent years much effort has been devoted to the characterization of nonequilibrium systems for which no general
theoretical framework is available.              Some notable progress has
been achieved in the study of rather generic properties of these systems, as for example the
entropy production and currents in nonequilibrium steady states \cite{Eva93,Gal95,Leb99,Ber07,Der07,Meh08},
the aging phenomena in systems relaxing towards steady states \cite{Hen10},
or the fluctuation properties of systems driven out of a steady state \cite{Jar97,Cro99,Cro00,Jar00,Hat01,Spe05,Har07}.
Interestingly, a large part of this progress is closely related to the discovery of various fluctuation and work theorems
that provide generic statements applicable to large classes of systems.

One of the best known theorems is Crooks relation for a system, initially in equilibrium, that is driven out of equilibrium
through a time-dependent process. Repetition of this forward process allows
to determine the probability distribution $P_F(W_d)$ of the dissipative work $W_d$. When comparing this
probability distribution with that obtained from the time-reversed process, called $P_R$, one observes the following
very simple relation
\[
P_F(\beta W_d)/P_R(-\beta W_d) = e^{\beta W_d}~,
\]
with $\beta = 1/(k_BT)$. This detailed fluctuation theorem is closely connected to other relations, notably
Jarzynski's work theorem. The latter is an integral fluctuation relation that relates for the special case of equilibrium initial and
final states the free energy difference $\Delta F$ to an average over all processes
leading from one state to the other:
\[
\left< e^{- \beta W} \right> = e^{- \beta \Delta F}~,
\]
where $W$ is the work done on the system.

Both the Jarzynski and the Crooks relations have been generalized to various other situations. Assuming microreversibility,
integral and detailed fluctuation theorems \cite{Hat01,Har07,Che06} have been derived for the
Hatano-Sasa entropy $\phi$, also called the driving entropy production, a quantity
that is closely related to the excess heat. For an equilibrium system
$\phi$ reduces to the dissipative work $W_d$ and the theorems become
identical to the Jarzynski and Crooks relations. In addition, an integral fluctuation theorem has also been formulated
for the housekeeping heat \cite{Spe05}. Finally, it was shown recently that under the same assumptions of
microreversibility integral and detailed fluctuation theorems also hold for the adiabatic and nonadiabatic
trajectory entropies \cite{Esp10}.

In absence of microscopic reversibility, as it is for example the case in reaction-diffusion systems, a detailed fluctuation theorem,
which is based on a forward path and the time-reversed path, does not hold \cite{footnote}.
However, recent studies of two of us
revealed that for the Hatano-Sasa entropy the deviations of the fluctuation ratio 
from a simple exponential
contain non-trivial information on the trajectories in configuration space \cite{Dor09,Dor09b,Dor10}.

In this work we revisit the fluctuation relations involving the Hatano-Sasa entropy when driving a system
out of a steady state. We confirm that an integral fluctuation theorem is always recovered, irrespective of
whether the processes are reversible or not. In the limit
of slow processes, where the time between changes of some external parameter is long enough for the system
to reach the steady state, a detailed fluctuation theorem prevails. Finally, we discuss many-state processes
that are irreversible and violate detailed balance and show that for large classes of these processes Crooks relation
for the Hatano-Sasa entropy remains valid. This remarkable result is obtained through a coarsening theorem
that states that some many-state
processes can be coarse-grained into systems with fewer states that have the same spectrum of $\phi$ and the
same probabilities for each $\phi$ as the original systems.

Our paper is organized as follows. Sections II and III focus on exact results on the realm of validity of
fluctuation relations involving the Hatano-Sasa entropy. Section II is mainly devoted to the case of slow processes,
whereas in Section III we discuss the coarse-graining procedure and its consequences for the detailed
fluctuation theorem in many-state processes. In Section IV we illustrate our main results through numerical studies
of selected systems. Finally, Section V gives our conclusions.

\section{Slow processes}

We consider in the following a stochastic dynamical process driven by an external parameter $\g$, and characterized by its internal
state $x(t)$~\cite{remark1}.  The system is taken through a series of changes, over a total period of time, $T$, whereupon the parameter $\g$ is varied from $\g=\g_0$ to $\g=\g_M$ in a series of $M+1$ steps, each lasting a time $\tau=T/(M+1)$.
We denote the {\em steady state} pdf of the system, under $\g=\g_n$, by $\rho(x,\g_n)$, and write it, more compactly, as
\[
\r(x,\g_n)\equiv \r_n(x)\,.
\]


Let $P_n(x'|x'')$ denote the transition probability, in the time interval $\tau$, from the initial state $x(t=0)=x''$ to the final state $x(t=\tau)=x'$, when the driving parameter is $\g=\g_n$.  The weight of the particular {\em path} $[x,\g]=(x_0,\g_0),(x_1,\g_1),\dots,(x_M,\g_M)$ is given by
\begin{equation}
\label{Wpath}
W([x,\g])=\r_0(x_0)\prod_{n=0}^{M-1}P_{n+1}(x_{n+1}|x_n)\,,
\end{equation}
and the average of a dynamical quantity, $Q([x])$, over all possible histories,  is given by
\[
\av{Q}=\int\cdots\int Q([x])W([x,\g])\,dx_0\,dx_1\cdots dx_M\,.
\]

Suppose that the initial pdf of the system, at time $t$, is $p(x,t)=\r_n(x)$, then the pdf  at time $t+\tau$ is
\begin{equation}
\label{rPr}
\int P_n(x|y)\r_n(y)\,dy=\r_n(x)\,,
\end{equation}
by the very definition of the steady state: Since the system is in the steady state to begin with,  it will remain there regardless of how long it evolves (assuming that the external parameters are held constant).  If the initial state is different, for example, $p(x,t)=\delta(x-x')$, the pdf after time $\tau$,
\[
p(x,t+\tau)=\int P_n(x|y)\delta(y-x')\,dy=P_n(x|x')\,,
\] 
is {\em not} necessarily equal to the steady state $\rho_n(x)$.  Note, however, that if $\tau\to\infty$ the system would eventually arrive at the steady
state, regardless of its initial condition:
\begin{equation}
\label{Pslow}
p(x,t+\tau)=\int P_n(x|y)\delta(y-x')\,dy=P_n(x|x')\to\r_n(x),\qquad{\rm as\ }\tau\to\infty\,.
\end{equation}
We call {\em slow process} a process where the system is allowed to reach the steady state after each incremental change in the external parameters (by letting $\tau\to\infty$).  Using (\ref{Pslow}), we see then that the weight of a path for a slow process assumes the much simpler form:
\begin{equation}
\label{Wslow}
W_{\rm slow}([x,\g])=\prod_{n=0}^{M}\r_n(x_n)\,;\qquad{\rm slow\ process.}
\end{equation}
In the following we explore the validity of different fluctuation relations within the setting just described.

\subsection{The integral fluctuation theorem, $\av{e^{-\phi}}=1$}

Define $\Delta\phi$, during one step, as
\[
\Delta\phi_n=\ln\r_n(x_n)-\ln\r_{n+1}(x_n)\,,
\]
such that the total change throughout the process is
\begin{equation} \label{eq:phi}
\phi=\sum_{n=0}^{M-1}\Delta\phi_n=\sum_{n=0}^{M-1}[\ln\r_n(x_n)-\ln\r_{n+1}(x_n)]\,.
\end{equation}
Using the path weight (\ref{Wpath}), we then have
\begin{equation}
\label{ephi}
\av{e^{-\phi}}=\int\cdots\int\r_0(x_0)\prod_{n=0}^{M-1}\frac{\r_{n+1}(x_n)}{\r_n(x_n)}P_{n+1}(x_{n+1}|x_n)\,dx_0\,dx_1\cdots dx_M=1\,.
\end{equation}
This can be seen most easily by integrating over the variables $x_0,x_1,\dots,x_M$ in this precise order, and
using the relation~(\ref{rPr}).  Thus, $\av{\exp(-\phi)}=1$.

It has to be noted that this result is valid for all processes, as we did not make any restricting assumptions.
Especially, the processes do not need to be reversible nor slow.

\subsection{The pdf $P(\phi)$ and the ratio $\pf(\phi)/\pr(-\phi)$}

The pdf of $\phi$ is given by
\begin{equation} \label{eq:Pphi}
P(\phi)=\int\cdots\int \delta\left(\phi-\sum_{n=0}^{M-1}[\ln\r_n(x_n)-\ln\r_{n+1}(x_n)]\right)W([x,\g])\,dx_0\,dx_1\cdots dx_M\,.
\end{equation}
Since the integrals involved are often hard to evaluate, we prefer working with the Fourier transform:
\begin{equation}
\label{eq:FPphi}
\F{P(\phi)}=\int_{-\infty}^{\infty}P(\phi)\,e^{-i\phi k}\,d\phi=\av{e^{-i\phi k}}\,.
\end{equation}

We now turn to the ratio of the pdf's for the {\em forward} and {\em backward} processes.  The forward process is the
process we have been considering all along.  The backward process, is the same process but where the order
of states is reversed.
It can be thought of as a forward process along the reversed path $[x,\g]=(x_M,\g_M),(x_{M-1},\g_{M-1}),\dots,(x_0,\g_0)$, obtained by making the substitution $n\to M-n$.   Note that in the backward process $(x_M,\g_M)$ is assumed
to be in the steady state, while $(x_0,\g_0)$ is generically {\em not} in the steady state.  Thus, the backward process
is not a perfect time-reversal of the forward process, but the reversal is only in the sequence of states.

%

Putting~(\ref{eq:Pphi}) and~(\ref{Wpath}) in~(\ref{eq:FPphi}) and carrying out the integration over $\phi$ first, we obtain
for the forward process
\[
\begin{split}
\F{P_F(\phi)}&=\int\cdots\int\rho_0(x_0)\prod_{n=0}^{M-1}\left\{\left[\frac{\rho_n(x_n)}{\rho_{n+1}(x_n)}\right]^{-ik}P_{n+1}(x_{n+1}|x_n)\right\}dx_0\cdots dx_M\\
&=\int\cdots\int\rho_0(x_0)^{1-ik}\prod_{n=1}^{M-1}\left\{\left[\frac{\rho_n(x_n)}{\rho_{n}(x_{n-1})}\right]^{-ik}P_{n}(x_{n}|x_{n-1})\right\}\rho_M(x_{M-1})^{ik}dx_0\cdots dx_{M-1}\,,
\end{split}
\]
where, for the second line, we have rearranged the product and integrated over $x_M$, using the fact that $\int P(y|x)\,dy=1$.
If the states of the system form a discrete set, $\{A,B,C,\dots\}$, the remaining integrals can be put in the form of a matrices product,
\begin{equation}
\F{P_F(\phi)}=\left(\rho_M(A)^{ik},\rho_M(B)^{ik},\dots\right){\bf F}_{M-1}{\bf F}_{M-2}\cdots {\bf F}_{1}\left(
\begin{array}{c}
\rho_0(A)^{-ik+1}\\
\rho_0(B)^{-ik+1}\\
\vdots
\end{array}
\right)\,,
\end{equation}
with 
\[
\left({\bf F}_n\right)_{XY}=P_n(X|Y)\frac{\rho_n(X)^{-ik}}{\rho_n(Y)^{-ik}}\,.
\]
Similarly, the transform of $P_R(-\phi)e^{\phi}$ can be written as
\begin{equation}
\label{FPR}
\F{P_R(-\phi)e^{\phi}}=\left(\rho_M(A)^{-ik+1},\rho_M(B)^{-ik+1},\dots\right){\bf R}_{1}{\bf R}_{2}\cdots {\bf R}_{M-1}\left(
\begin{array}{c}
\rho_0(A)^{ik}\\
\rho_0(B)^{ik}\\
\vdots
\end{array}
\right)\,,
\end{equation}
where now
\[
\left({\bf R}_n\right)_{XY}=P_n(X|Y)\frac{\rho_n(Y)^{-ik+1}}{\rho_n(X)^{-ik+1}}\,.
\]
Upon taking the transpose of the rhs of~(\ref{FPR}) it is seen that $\F{P_R(-\phi)e^{\phi}}=\F{P_F(\phi)}$, provided
that ${\bf R}_n^T={\bf F}_n$.  But ${\bf R}_n^T={\bf F}_n$ if and only if
$P_n(X|Y)\rho_n(Y)=P_n(Y|X)\rho_n(X)$.  This last relation is satisfied in all {\em equilibrium} processes, where it is known as {\em detailed balance}.

In the limit of slow processes, 
$P(x|y)\to\rho(y)$, hence $P(x|y)\rho(y)=\rho(x)\rho(y)=\rho(y)\rho(x)=P(y|x)\rho(x)$.  Therefore 
${\bf R}_n^T={\bf F}_n$ (even for nonequilibrium systems) and the Crooks relation $P_F(\phi)/P_R(-\phi)=e^{\phi}$ is satisfied.

\section{Coarse-graining of many-state processes}
As we discuss in the following the validity of Crooks relation is much larger than what one would expect naively. In fact,
there are large classes of nonequilibrium systems that obey this relation, and this
despite their irreversibility and the violation of detailed balance.

Let us first consider a system with only two states, $A$ and $B$.  Unless there is at least one transition (say, from $A$ to $B$), 
the system is trivial and does not evolve at all.  However, with only $A\to B$ the steady state of the system is $\rho(A)=0$
and the generalized entropy $\phi$ is not well defined.  We therefore need consider only systems that include
both transitions, $A\to B$ and $B\to A$.   Such systems, however, have an equilibrium state that obeys detailed balance,
and therefore the Crooks relation is satisfied for all (relevant) two-state systems.

Generally, irreversible processes with more than two states are not expected to obey the Crooks relation.  We now identify
a wide class of such systems that does satisfy the Crooks relation, despite their patent violation of detailed balance.
These are characterized by the following theorem:

\medskip\noindent
Any system with states $A_1,A_2,\dots;B_1,B_2,\dots;C_1,C_2,\dots;\dots$, such that all the transition rates $A_i\to X_{i'}$ are of the form $\omega(A_i,X_{i'})\alpha$, all the rates $B_j\to Y_{j'}$ are of the form $\omega(B_j,Y_{j'})\beta$, the rates $C_k\to Z_{k'}$ are $\omega(C_k,Z_{k'})\gamma$, etc., where all the $\omega$'s are {\em constants} and $\alpha,\beta,\gamma,\dots$
are driving parameters that are varied arbitrarily through the process' duration,  can be ``coarse-grained" into a system with states $A,B,C,\dots$, where the ``super-states" $X=\bigcup_iX_i$ are an aggregate of the states $X_i$.  The pdf's $P(\phi)$ of the original system and the coarse-grained system are exactly the same.
\medskip

For the special case where there are only two types of states, $A_i$ and $B_j$, the system is equivalent to
a two-state coarse-grained system and the Crooks relation is obeyed!

To prove the coarsening theorem we write down the rate equations,
\begin{equation}
\begin{split}
\dot\rho(A_i)&=-\sum_{X_k}{\rm rate}(A_i\to X_k)\rho(A_i)+\sum_{Y_k}{\rm rate}(Y_k\to A_i)\rho(Y_k)\,,\\
\dot\rho(B_j)&=-\sum_{X_k}{\rm rate}(B_j\to X_k)\rho(B_j)+\sum_{Y_k}{\rm rate}(Y_k\to B_j)\rho(Y_k)\,,\\
&\>\>\vdots
\end{split}
\end{equation}
where $\rho(X_i)$ now denotes the time-dependent probability for finding the system in state $X_i$ and the overdot denotes
time differentiation.  Using  ${\rm rate}(X_i\to Y_j)=\xi\omega(X_i\to Y_j)$, as postulated by the theorem's premise, and writing $\rho(X_i)=\xi^{-1} r(X_i)$
for all the various $X$-species, the equations become:
\begin{equation}
\begin{split}
\dot \rho(A_i)&=-\sum_{X_k}\omega(A_i\to X_k)r(A_i)+\sum_{Y_k}\omega(Y_k\to A_i)r(Y_k)\,,\\
\dot \rho(B_j)&=-\sum_{X_k}\omega(B_j\to X_k)r(B_j)+\sum_{Y_k}\omega(Y_k\to B_j)r(Y_k)\,,\\
&\>\>\vdots
\end{split}
\end{equation}
We can now see that in the steady state, when the left side of the rate equations is set to 0, the $r(X_i)$ satisfy
a homogeneous linear system of equations with constant coefficients and therefore they are constant (independent
of the driving fields $\alpha,\beta,\dots$).  Then, for states of type $X$, the {\em steady-state} probability is 
\begin{equation}
\label{rho_r}
\rho(X_i)=c(\alpha,\beta,\dots)\xi^{-1}r(X_i)\,,
\end{equation} 
where $c$ is a normalization constant, determined by $\sum_{X_i}\rho(X_i)=1$.  Note that while $c$ is a function of the driving fields, it has the very same value for {\em all} states (of all types).

Recall now the expression for $\phi$:
\[
\phi = \ln\left(\prod_{n=0}^{M-1}\frac{ \rho_n(x_n)}{\rho_{n+1}(x_n)}\right)\,.
\]
The important thing is that in each $\rho_n(x_n)/\rho_{n+1}(x_n)$-factor there appears the same state $x_n$ (both
in the numerator and the denominator), therefore, from~(\ref{rho_r}),  each such term depends only on the type of  the state $x_n$, but
not on which particular one.  Thus, for example, $\rho_n(A_i)/\rho_{n+1}(A_i)=c_n\alpha_{n+1}/c_{n+1}\alpha_n$ is independent of $i$.  It follows
that the {\em spectrum} of $\phi$ (the possible values it attains) depends only on the order that the different {\em types} of
states are met: The paths $X_i\to Y_j\to Z_k\to W_l\to\cdots$ have the same value of entropy, $\phi=\phi'$, for all $i,j,k,l,\dots$ 
~~The probability for having this particular value of entropy, $P(\phi')$, is given by the sum of all the pertinent path weights:
\[
P(\phi')=\sum_{i,j,k,l,\dots}\rho_0(X_i)P_1(Y_j|X_i)P_2(Z_k|Y_j)P_3(W_l|Z_k)\cdots\,.
\]

Consider now the coarse-grained system, where we associate all the states of one type with a single ``super-state,"
$X=\bigcup_iX_i$, ($X=A,B,C,\dots$).  The probability that the system is initially in super-state $X$ is
$\rho_0(X)=\sum_i\rho_0(X_i)$,
while 
\[
\sum_{i,j,k,l}\rho_0(X_i)P_1(Y_j|X_i)P_2(Z_k|Y_j)P_3(W_l|Z_k)=\rho_0(X)P_1(Y|X)P_2(Z|Y)P_3(W|Z)
\]
denotes the probability for being at super-state $W$ at step 3, having followed the path $X\to Y\to Z\to W$ (and likewise for additional steps).
The conclusion that the coarse-grained system has the same spectrum of $\phi$ and the same probabilities for each $\phi$
as the original system follows immediately.

\section{Numerical results}
In this Section we  illustrate our results  through the numerical study of some nonequilibrium
reaction networks. The systems discussed in the following can be in various states, called $A, B, C, \cdots$, and the
passage from one state to another takes place with a certain reaction rate.

\subsection{Three states}
\begin{figure}[h]
\centerline{\epsfxsize=4.00in\ \epsfbox{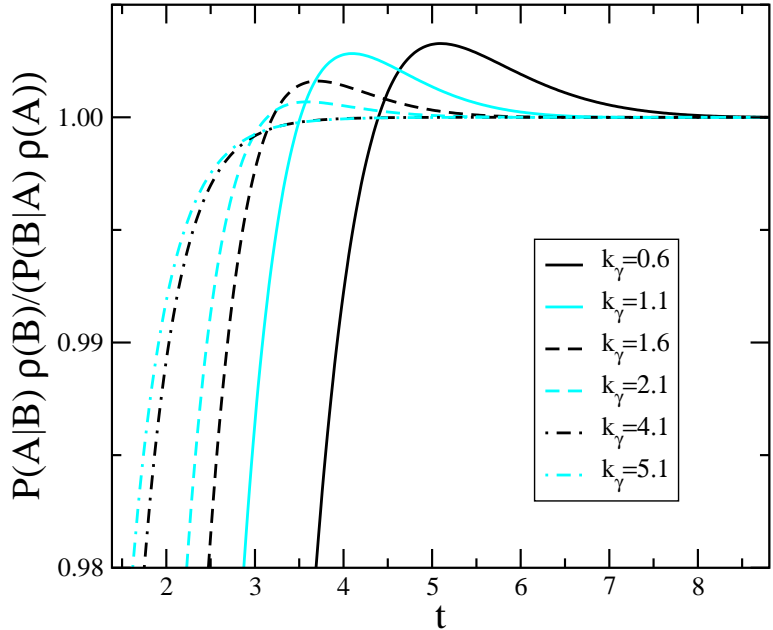}}
\caption{(Color online) Detailed balance ratio for the three-states cyclic model as a function of time for various
values of $k_\gamma$, with $k_\alpha = k_\beta = 1$. Note the non-monotonic approach to stationarity when
$k_\gamma < 4$.
}
\label{fig1}
\end{figure}

Let us start with the simple case of a three-state system with the cyclic reaction scheme
\begin{eqnarray}
A & \overset{k_\alpha}{\rightarrow} & B \nonumber \\
B & \overset{k_\beta}{\rightarrow} & C \nonumber \\
C & \overset{k_\gamma}{\rightarrow} & A \nonumber 
\end{eqnarray}
The rate equations are given by
\begin{eqnarray}
\dot{\rho}(A) & = & - k_\alpha \rho(A) + k_\gamma \rho(C) \nonumber \\
\dot{\rho}(B) & = & - k_\beta \rho(B) + k_\alpha \rho(A) \nonumber \\
\dot{\rho}(C) & = & - k_\gamma \rho(C) + k_\beta \rho(B) \nonumber
\end{eqnarray}
with $\rho(A) + \rho(B) + \rho(C) = 1$. The steady-state probabilities are readily shown 
to be $\rho_0(B)= \frac{k_\alpha}{k_\beta} \rho_0(A)$, $\rho_0(C)= \frac{k_\alpha}{k_\gamma} \rho_0(A)$, with
\[
\rho_0(A)= \frac{k_\beta k_\gamma}{k_\beta k_\gamma + k_\gamma k_\alpha + k_\alpha k_\beta}~.
\]
The rate equations can be integrated straightforwardly, yielding expressions
for the time-dependent transition probabilities.  For example,  $P(Y|X)$ is obtained by using the initial condition $\rho(Z,t=0)=\delta_{Z,X}$  and integrating to time $t=\tau$.
At constant values of the reaction rates, the system rapidly evolves towards the steady state. While detailed balance does not generally hold, the relation $P(X|Y)\rho(Y)=P(Y|X)\rho(X)$ gets fulfilled as $t\to\infty$.  This is illustrated
in Fig.~\ref{fig1}, where we plot the detailed balance ratio 
$ \frac{P(A|B) \rho(B)}{P(B|A) \rho(A)}$ as a function of time for fixed values $k_\alpha = k_\beta = 1$ and various
values of $k_\gamma$. 
Amusingly, the approach to stationarity is non-monotonous for $k_\gamma < 4$.

Our focus in this paper is on situations where the reaction rates are not constant but are changed in time.
The protocol that we use is the same as that discussed in the previous Sections: some (or all) reaction
rates are changed $M+1$ times over a total period of time $T$, such that between two changes the rates are kept 
constant for a time interval of length $\tau = T/(M+1)$. The transition probabilities between the different states
can be found analytically for the simpler cases. For the more complex situations, we obtain these transition
probabilities by numerically integrating the rate equations. Once we have the transition probabilities, we can compute
the probability distribution of $\phi$ using Eqs. (\ref{eq:Pphi}) and (\ref{Wpath}).

\begin{figure}[h]
\centerline{\epsfxsize=4.50in\ \epsfbox{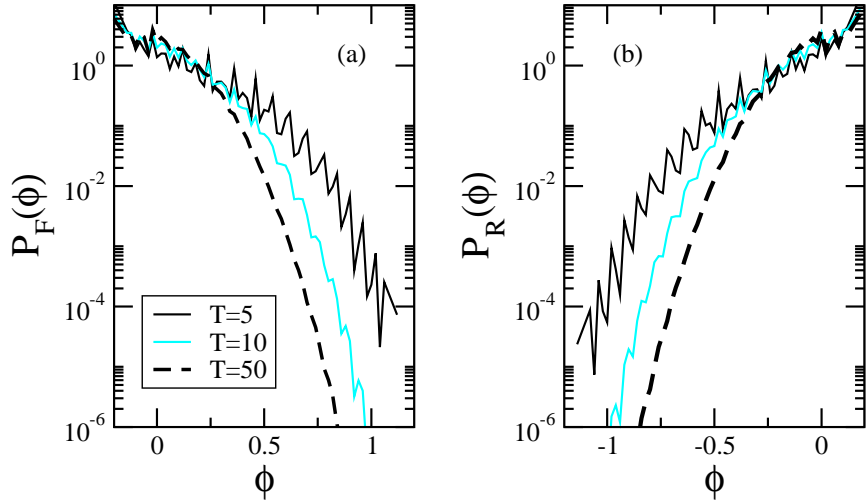}}
\caption{(Color online) Probability distributions for the cyclic three-state model for (a) the forward process and (b)
the reversed process. The different curves correspond to different total times $T$.
In the forward process the rate $k_\gamma$ is changed from 1 to 4 in $M = 15$ steps, with $k_\alpha = 0.5$ and $k_\beta =1$.
}
\label{fig2}
\end{figure}

As an example, we show in Fig.~\ref{fig2} probability distributions obtained for our three-state model
where one rate, namely $k_\gamma$, is increased from $k_\gamma = 1$ to $k_\gamma = 4$ in $M = 15$ steps, see Fig.~\ref{fig2}a.
The probability distribution for this forward process is denoted by $P_F(\phi)$ and the different curves
correspond to different total times $T$. In Fig. \ref{fig2}b we show the 
probability distribution $P_R(\phi)$ for the reversed process where we start with $k_\gamma = 4$ and decrease
that rate to $k_\gamma = 1$ in the same number of steps. 

Looking at these probability distributions, the first thing one notices is their irregular structures for
small times $T$, characterized by pronounced peaks. These peaks become less and less prominent when $T$ increases,
yielding a smooth distribution in the long-time limit. In addition, the distributions rapidly converge to a limiting
curve when $T$ increases, and no notable changes in the shapes of the distributions are measured when $T$ exceeds 50.

\begin{figure}[h]
\centerline{\epsfxsize=4.50in\ \epsfbox{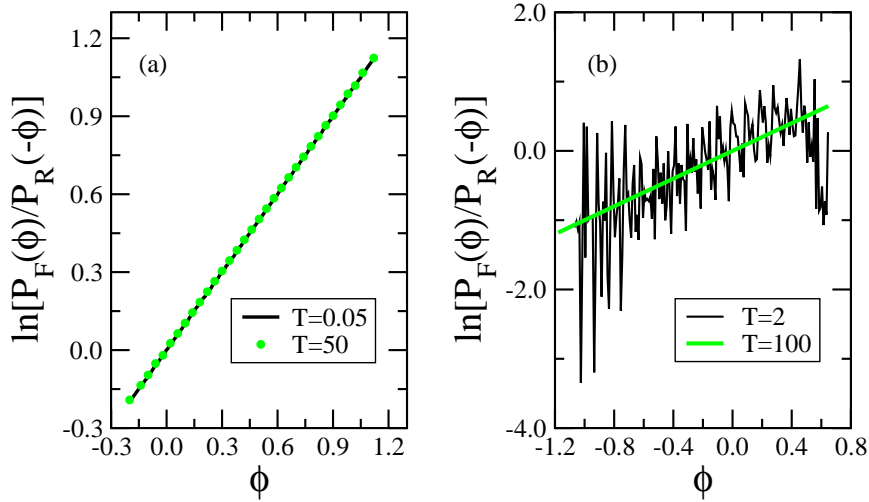}}
\caption{(Color online) Fluctuation ratios for the three-state cyclic model. In (a) only the rate $k_\gamma$ is changed from
$k_\gamma = 1$ to $k_\gamma = 4$ in $M =15$ steps, the other rates being kept fixed, with $k_\alpha = 0.5$ and $k_\beta = 1$.
In (b) all three rates are varied independently: $k_\alpha$ from 2.5 to 1.5, $k_\beta$ from 0.75 to 2.25, and $k_\gamma$ from
1.1 to 4.1, again in $M =15$ steps. For both cases the fluctuation ratios for two different times $T$ are shown.
}
\label{fig3}
\end{figure}

In order to verify the predictions of the previous Sections, we vary the reaction rates in different ways and
compute the fluctuation ratio $P_F(\phi)/P_R(-\phi)$. Some of our results are summarized in Fig.~\ref{fig3}.
Whereas in Fig.~\ref{fig3}a only the rate $k_\gamma$ is changed, the rates $k_\alpha$ and $k_\beta$ being constant,
in Fig.~\ref{fig3}b we vary all three rates in an independent way. The first case can be viewed as a simple example
of a system that is equivalent to a two-state coarse-grained system (the states $A$ and $B$ can be coarse-grained into a single state), see Section~III.
 Consequently, we expect Crooks relation
\[
P_F(\phi)/P_R(-\phi) = e^{\phi}
\]
to be fulfilled, independently of the length of the time interval. As shown in Fig.~\ref{fig3}a, this is indeed
the case. When all rates are varied independently, the system can not be replaced by an equivalent 
coarse-grained two-state system, and Crooks relation does not hold, as shown in Fig.~\ref{fig3}b for $T = 2$.
For very large $T$, however, the process becomes slow, and Crooks relation is then again recovered, in
accordance with our discussion in Section II. 

\subsection{Six states}

\begin{figure}[h]
\centerline{\epsfxsize=3.00in\ \epsfbox{
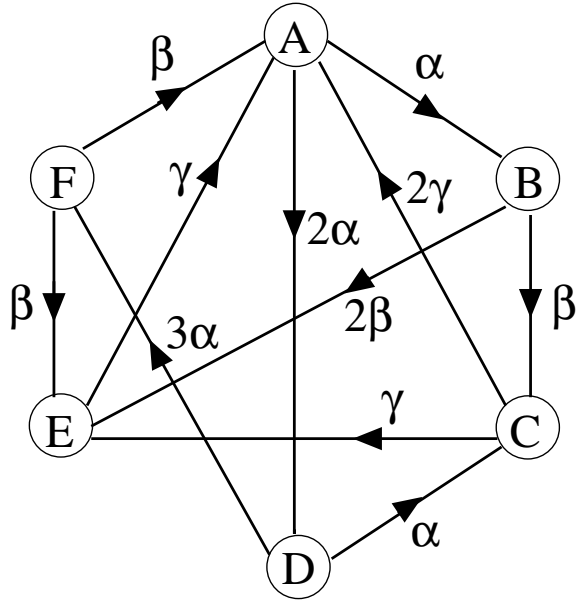}}
\caption{A six-state process that can be coarse-grained into an equivalent three-state process when
$\alpha$, $\beta$, and $\gamma$ are varied independently.
}
\label{fig4}
\end{figure}

In order to fully appreciate the coarse-graining procedure and the generality of our results, we study
in the following the rather complex six-state system given in Fig.~\ref{fig4}.
Inspection of that figure reveals that the processes that lead out of the states $A$ and $D$ depend only on
the rate $\alpha$. Similarly, processes out of the states $B$ and $F$ depend only on $\beta$, and out of $C$ and $E$, only
on $\gamma$. Therefore, if $\alpha$, $\gamma$ and $\beta$ are varied independently, this six-state system should be equivalent to 
a three-state system, where the new states result from the union of pairs of the original states:
$\overline{AD} = A \cup D$, $\overline{BF} = B \cup F$, $\overline{CE} = C \cup E$.
If, in addition, we set $\beta = \gamma$, for example, we have only two independent variables, and our six-state process
should be equivalent to a two-state system. Whereas in the former case the detailed fluctuation theorem should be broken,
in the latter case we should recover Crooks relation even for our six-state system.
%
\begin{figure}[h]
\centerline{\epsfxsize=4.50in\ \epsfbox{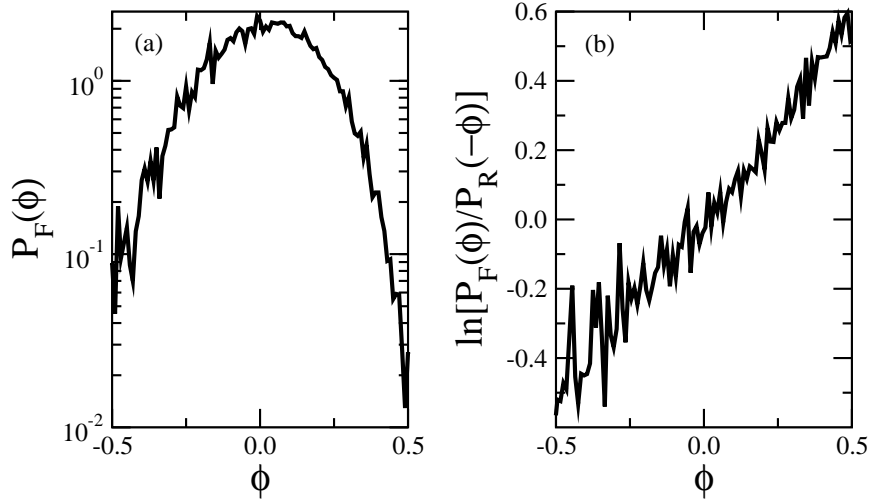}}
\caption{(Color online) (a) Probability distribution of $\phi$ and (b) fluctuation ratio for the six-state
model when $\alpha$ is changed from 1 to 3 and $\beta$ from 1 to 2,
whereas $\gamma = 2$ is kept constant. The changes are done in $M = 8$ steps, with $T = 5$. For this choice of the parameters,
the model does not have an equivalent two-state model and the Crooks relation is not fulfilled.
}
\label{fig5}
\end{figure}

\begin{figure}[h]
\centerline{\epsfxsize=4.50in\ \epsfbox{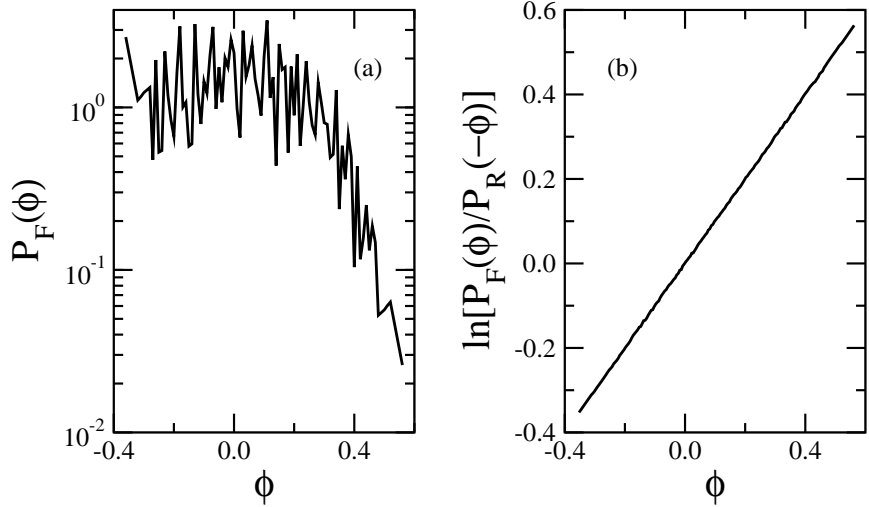}}
\caption{(Color online) (a) Probability distribution of $\phi$ and (b) fluctuation ratio for the six-state 
model when $\alpha$ is changed from 1 to 5 whereas $\beta = \gamma$ is
changed from 2 to 4. The changes are done in $M = 8$ steps, with $T = 5$. For this choice of the parameters, the model
is equivalent to a two-state model and the Crooks relation is fulfilled.
}
\label{fig6}
\end{figure}

We carefully checked that our original six-state model has indeed the same spectrum and probabilities for $\phi$
as the coarse-grained systems, in accordance 
with the theorem proven in Section~III. In Fig.~\ref{fig5} we vary the three rates
$\alpha$, $\beta$, and $\gamma$ independently. Consequently, the six-state model can not be reduced to an
equivalent two-state model and the Crooks relation is not fulfilled. This is different for the case
shown in Fig.~\ref{fig6}, where $\beta = \gamma$ throughout the whole process. The six-state model can 
then be reduced to a coarse-grained two-state model and the Crooks relation is fulfilled.

\section{Conclusion}
Understanding the properties of systems far from equilibrium is of great importance in many fields in physics.
However, the challenges encountered in that endeavor remain extraordinary. In that context, the recent formulation
of various fluctuation theorems, applicable either to systems in their steady states or to systems driven
out of stationarity, constitutes an important development, as they allow a certain characterization 
and classification of various far from
equilibrium systems.
For that reason it is important to further probe these theorems, in order to better understand their range of
applicability.

In this work we have focused on the Hatano-Sasa entropy $\phi$, a  quantity which reduces to the dissipative work in the special
case of a system driven out of equilibrium, and that remains well defined even in the absence of microscopic reversibility.
This entropy fulfills both an integral and a detailed fluctuation theorem for slow systems, i.e., systems which
reach the steady state after each change in the external parameters. Most importantly, we show that for large classes
of processes with many states one can find equivalent processes with fewer states that have the same spectrum
of $\phi$ and the same probabilities for each $\phi$. In the special case that this coarse-grained system is a two-state system,
the original many-state process verifies Crooks relation, and this even when detailed balance is
broken due to the absence of microscopic reversibility.

Our results show that Crooks detailed fluctuation theorem is much more widely applicable than thought previously,
provided that the Hatano-Sasa entropy is used for characterizing the system. This opens the possibility to describe
through fluctuation theorems wide classes of systems that have not been considered in that context previously.

%

\begin{acknowledgments}
This work was supported by the US National
Science Foundation through DMR-0904999.
\end{acknowledgments}


\begin{thebibliography}{99}
\bibitem{footnote} In absence of microreversibility fluctuation theorems can still be formulated in terms
of posterior probabilities \cite{Ohk09}.
\bibitem{remark1} The driving force can consist of more than one field.  We denote all these fields, collectively, with the parameter $\g$.  Similar remarks apply to the system's state, $x$, that may involve more than one component.
\end{thebibliography}
\end{document}